\documentclass[aps,prb,a4paper,twocolumn,showpacs,amssymb,amsfonts,amsmath]{revtex4-1}
\usepackage{bm,graphicx,verbatim}
\usepackage[dvips]{hyperref}
\renewcommand{\vec}[1]{{\mathbf #1}}
\newcommand{\be}{\begin{equation}}
\newcommand{\ee}{\end{equation}}
\newcommand{\Neel}{{N\'{e}el}}
\newcommand{\ZnCrO}{{$\rm ZnCr_2O_4$}}
\newcommand{\MgCrO}{{$\rm MgCr_2O_4$}}
\newcommand{\CdCrO}{{$\rm CdCr_2O_4$}}
\newcommand{\CdFeO}{{$\rm CdFe_2O_4$}}
\newcommand{\ZnFeO}{{$\rm ZnFe_2O_4$}}
\newcommand{\DyTiO}{{$\rm Dy_2Ti_2O_7$}}
\newcommand{\HoTiO}{{$\rm Ho_2Ti_2O_7$}}

\newcommand{\refeq}[1]{{Eq.~\eqref{#1}}}

\begin{document}
\title{Absent pinch points and emergent clusters: further neighbour interactions in the pyrochlore Heisenberg antiferromagnet}
\author{P.\ H.\ Conlon}
\email{conlon@thphys.ox.ac.uk}
\author{J.\ T.\ Chalker}
\affiliation{Theoretical Physics, Oxford University, 1 Keble Road, Oxford, OX1 3NP, United Kingdom}
\date{May 14, 2010}
\begin{abstract}
We discuss the origin of spin correlations observed in neutron scattering experiments on the paramagnetic phase of a number of frustrated spinel compounds, most notably $\rm ZnCr_2O_4$. These correlations are striking for two reasons. First, they have been interpreted as evidence for the formation of weakly interacting hexagonal clusters of spins. Second, they are very different from those calculated for the nearest neighbour Heisenberg pyrochlore antiferromagnet, in which Coulomb phase correlations generate sharp scattering features known as pinch points. Using large-$n$ calculations and Monte Carlo simulations, we show that very weak further neighbour exchange interactions can account for both the apparent formation of clusters and the suppression of pinch points.
\end{abstract}

\maketitle
\section{Introduction}
A central reason for interest in geometrically frustrated magnets is that they are systems which can support strong correlations without developing long range order. This cooperative paramagnetic behaviour appears at temperatures small compared to the main interaction scale, which is characterised by the magnitude of the Curie-Weiss constant $\Theta_{\rm CW}$. The absence of long range order at low temperature can be understood in the framework of classical models with nearest neighbour interactions, as a consequence of macroscopic ground state degeneracy.
\cite{[For introductions see: ] Review1, *Review2}

Results from recent neutron scattering studies of some important examples of geometrically frustrated magnets present a paradox, and our aim in this paper is to offer a resolution of the paradox. More specifically, experiments on a number of spinel compounds, including in particular  \ZnCrO, have been interpreted as revealing the formation of small, independent clusters of spins\cite{LeeBroholm02,TomiyasuSuzuki08,ChungMatsuda05,KamazawaParkLee04}. In contrast, the minimal theoretical model which is assumed to describe these systems, the nearest neighbour Heisenberg pyrochlore antiferromagnet, is notable for a quite different feature: pinch point scattering, indicative of algebraic correlations. We show in this paper that further neighbour interactions strongly influence scattering in the cooperative paramagnetic phase. Very weak further neighbour interactions of the appropriate sign (ferromagnetic second neighbour, or antiferromagnetic third neighbour) are sufficient to suppress pinch-point scattering and generate quite accurately the short-range correlations that have been interpreted in terms of independent spin clusters.

\begin{figure}
\includegraphics{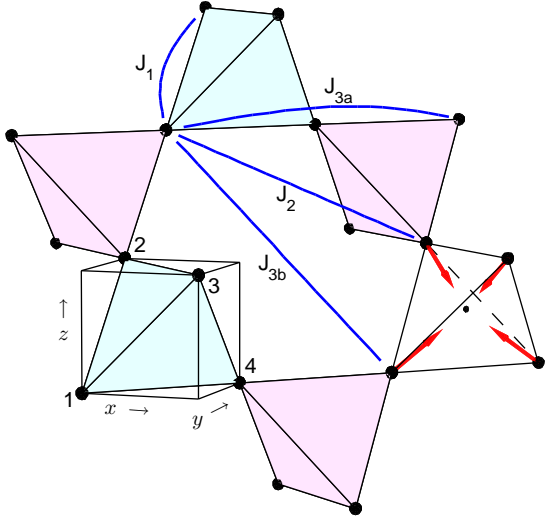}
\caption{\label{fig:lattice} (color online) Pyrochlore lattice, illustrating the first, second and third neighbour couplings. There are two inequivalent third neighbour interactions, but for $J_{3a}=J_{3b}$, we use $J_3$ to denote both. Numbers indicate sublattice labeling. The centres of the tetrahedra form a diamond lattice and the red arrows show the choice made to orient the diamond lattice bonds. }
\end{figure}

The spinel compounds, $AB_2O_4$, have a structure in which the $B$-sites form a three-dimensional network of corner-sharing tetrahedra, known as a pyrochlore lattice (Fig.~\ref{fig:lattice}). The same lattice is realized by both $A$ and $B$ sites in the pyrochlore compounds $A_2B_2O_7$, and in the Laves phase compounds with formula $AB_2$. Evidence for spin clusters has been reported in the chromites, $A\rm Cr_2O_4$ with non-magnetic $A$ = Zn, Mg and Cd as well as in the ferrite \CdFeO\ (Refs.~\onlinecite{LeeBroholm02,TomiyasuSuzuki08,ChungMatsuda05,KamazawaParkLee04} respectively). In all cases, the strong similarity between experimentally measured spin correlations and the form factor for independent hexagonal clusters leads to an interpretation in terms of independent spin clusters. Results for $\rm Y(Sc)Mn_2$ have broadly similar features\cite{BallouLelievre96}.
Cluster like correlations are also observed in the spin ice compound \DyTiO: here it has been argued from comparison with simulations that they arise from further neighbour interactions\cite{Yavorskii08}.
Spin clusters are not universally observed in frustrated pyrochlore magnets, however. Among the spin ice compounds, compelling evidence for pinch point scattering is seen in \HoTiO\ \cite{FennellDeenWildes09}.  For a recent review of properties of the frustrated spinels, see Ref.~\onlinecite{LeeTakagiLouca10}.

The classical nearest neighbour antiferromagnetic Heisenberg model on this lattice (which we refer to as the {\it NN model}) is now well understood theoretically\cite{Reimers92,MoessnerChalker98prl,MoessnerChalker98prb}. The model has no phase transition and remains in the paramagnetic phase down to temperature $T=0$. The low temperature regime is nevertheless strongly correlated. The nearest neighbour interaction energy is minimised by any spin configuration for which the magnetization on every tetrahedron vanishes, giving a ground state manifold with a macroscopic number of degrees of freedom. While states on this manifold are weighted differently by thermal fluctuations, order-by-disorder effects are too weak to induce a N\'eel state. Simulations of neutron scattering show sharp features in the structure factor\cite{ZinkinHarrisZeiske97}. 
Known as pinch points, a clear understanding of their origin comes from the recognition that ground states of the NN model can be mapped onto configurations of a solenoidal lattice vector field\cite{IsakovGregor04,Henley05} (this mapping will be reviewed in Sec.~\ref{sec:sub:longwavelength}) whose long wavelength fluctuations are governed by an entropic Maxwell action. The pinch points in scattering are a consequence of dipolar correlations of the solenoidal field. (For a recent review of models with such a Coulomb phase, see Ref.~\onlinecite{Henley09}.) Although algebraic at $T=0$, at finite temperature spin correlations decay exponentially beyond a correlation length $\xi_{\parallel}$, and pinch points acquire a width $\sim \xi_{\parallel}^{-1}$. For spin ice, the correlation length diverges exponentially as $T\rightarrow0$ due to the finite energy cost for violations of the ice rules, but for the NN model with continuous degrees of freedom, only as $T^{-1/2}$. Since the NN model unambiguously gives rise to pinch points in scattering when $T\ll\theta_{CW}$, with thermal broadening at finite temperature, their absence in experiments on the paramagnetic phase of the frustrated spinels demands an explanation which goes beyond the NN model.

Here we extend the NN model by including further neighbour exchange terms (see Fig.~\ref{fig:lattice} for definitions) and consider their effect on paramagnetic scattering. Further neighbour terms, both ferromagnetic (FM) and antiferromagnetic (AFM), will in general have two effects: (i) to change scattering in the paramagnetic phase; (ii) to induce ordering at a low $T$. With regard to ordering transitions out of the low temperature paramagnetic phase, the role of further neighbour exchange has been considered before: from a mean field perspective, \textcite{ReimersBerlinskyShi91} examined possible ordered states with further neighbour exchange; and following simulations by \textcite{Tsuneishi07}, a detailed study by \textcite{ChernMoessnerTchernyshyov08} of the phase diagram with a weak $J_2$ confirmed the mean field predictions. There has been less work to pin down the effect of further neighbour exchange on paramagnetic scattering in Heisenberg models, although we note that an analysis of neutron scattering data using third neighbours has been done for the frustrated spinel \ZnFeO, which has ferromagnetic nearest neighbour interactions\cite{YamadaKamazawa02}. In spin ice, \textcite{Yavorskii08} have shown with extensive simulations that further neighbour interactions can promote cluster-like scattering.

Our main results are as follows: first, further neighbour exchange terms of the right sign (FM $J_2$ or AFM $J_3$) can be effective both at suppressing pinch point scattering intensity and at broadening pinch points; and second, within the manifold of NN ground states, they favour states with cluster-like correlations. These effects are noticeable even at temperatures two orders of magnitude larger than the temperature of the transition that the additional interactions induce. Using the language of the Coulomb phase description, these effects are summarised in part by the fact that further neighbour terms renormalise the Coulomb phase coupling constant.

A visual impression of our findings in relation to experimental reports of spin clusters is given by comparing figures that show the computed diffuse scattering in two cases. The first case is the nearest neighbour model, for which pinch points, broadened by finite temperature, are apparent in Figs. \ref{fig:afmj3correlations} (a) and  \ref{fig:afmj3correlations} (b). The second case is a model with additional weak further neighbour interactions, for which behaviour is shown in Figs. \ref{fig:afmj3correlations} (c) and \ref{fig:afmj3correlations} (d). In this case scattering at the positions of pinch points is heavily suppressed. In its place, rings of scattering intensity are formed of the kind that have been interpreted as indications of spin cluster formation in Refs.~\onlinecite{LeeBroholm02,ChungMatsuda05,KamazawaParkLee04,TomiyasuSuzuki08}.

The remainder of this paper is organised as follows. In Section~\ref{sec:modelandtheory} we present the model and detail the main analytic tool used, a self-consistent Gaussian approximation which is exact  for $n$-component spins in the large-$n$ limit. Here we also review the mapping to lattice vector fields, which facilitates a more straightforward discussion of the long wavelength physics in the vicinity of pinch points. In Sec.~\ref{sec:fmj2afmj3} we consider the effect of a FM $J_2$ or an AFM $J_3$ which are of the sign required to induce cluster-like scattering and suppress pinch points. In Sec.~\ref{sec:afmj2fmj3}, we summarise the effects of AFM $J_2$ or FM $J_3$, and we look briefly at other combinations in Sec.~\ref{sec:generalcouplings}. In Sec.~\ref{sec:dynamics} we discuss dynamics, describing how we expect the effects of further neighbour interactions to be more conspicuous in quasi-elastic scattering than in static correlations. We consider estimates of further neighbour exchange for relevant compounds in Sec.~\ref{sec:exchangeestimates}. We summarise our results in Sec.~\ref{sec:discussion}. Further details of various calculations can be found in Ref.~\onlinecite{ConlonThesis}.

\section{Model and theoretical approach}\label{sec:modelandtheory}
Our focus is on the low temperature behaviour in the paramagnetic phase of the classical isotropic Heisenberg model with Hamiltonian
\be\label{eq:fullhamiltonian}
H = \frac{1}{2}\sum_{ij} J_{ij} \vec{S}_i.\vec{S}_j
\ee
where the degrees of freedom are unit length, three-component vectors residing on the sites of the pyrochlore lattice and where the dominant interaction is nearest neighbour antiferromagnetic (see Fig.~\ref{fig:lattice}). It is known that behaviour of the nearest neighbour Heisenberg model is very well approximated by the large-$n$ limit of its generalisation to $n$-component spins\cite{CanalsGaranin01,IsakovGregor04}.


We also use this large-$n$ limit. Through its equivalence to a spherical model\cite{Stanley68}, we present our method in the language of a self-consistent Gaussian approximation (SCGA). The validity of this approach is confirmed by the excellent agreement with Monte Carlo (MC) simulations on the Heisenberg model, \refeq{eq:fullhamiltonian}, both with and without further neighbour terms -- results from Monte Carlo simulations appear as points in figures and results from the following analysis as solid lines. Our simulations were performed using the Metropolis algorithm for systems of 2048 spins, and 16384 spins. The longest runs, required for the lowest temperatures, were $10^8$ MC steps per spin. In general though, run lengths were chosen in order that the statistical errors are smaller than the symbol size in the plots.
\subsection{Self-consistent Gaussian approximation}\label{sec:sub:scga}

\begin{figure}
\includegraphics[width=8.5cm]{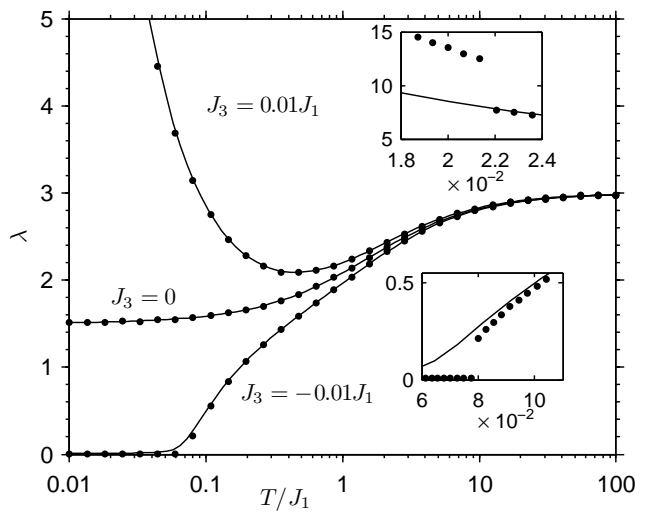}
\caption{Dependence of flux stiffness $\lambda$ on temperature, showing influence of weak further neighbour interactions. Solid lines: solutions of \refeq{eq:scgacondition} for the cases $J_3=\pm 0.01J_1$ and $J_3=0$, all with $J_2=0$. Points: results from simulations on 2048 spins interpreted via \refeq{eq:globalfluxfluctuations}. The effects of third neighbour interactions are clearly noticeable even at temperatures two orders of magnitude larger than $J_3$. Insets: detailed behaviour at low temperature; axes as in main panel. First order phase transitions occur in simulations for both signs of $J_3$.\label{fig:scgasolution}}
\end{figure}

We replace fixed length spins $\vec{S}_i$ with soft ones, denoting the magnitude of one component by $s_i$.
Soft spin configurations are weighted by $e^{-\beta {\cal H}}$ where
\be\label{eq:scgahamiltonian}
\beta {\cal H} = \frac{1}{2}\sum_{ij}(\lambda \delta_{ij} + \beta \sum_{n}J_n V^{(n)}_{ij})s_is_j\,.
\ee
The value of the Lagrange multiplier $\lambda$ is chosen to ensure $\langle s_i^2 \rangle = 1/3$ as required for a single component of a unit length three-component vector in a paramagnetic phase. The $m$-th neighbour interaction matrices $V^{(m)}_{ij}$ are related to the respective adjacency matrices by addition of multiples of the identity in the way described in Appendix~\ref{app:matrices} and these (seemingly arbitrary) choices of the diagonal entries are dictated by our subsequent interpretation of $\lambda$ as the stiffness of a Coulomb-phase action. The multiplier $\lambda$ is determined self-consistently by the condition on the average spin length
\begin{equation*}
\frac{1}{3} = \frac{1}{4N} {\rm Tr} \left[\lambda {\mathsf I} + \beta \sum_nJ_n {\mathsf V}^{(n)}\right]^{-1}
\end{equation*}
where $4N$ is the total number of sites. The translational symmetry of the system ensures that the interactions are block-diagonal in reciprocal space (see Appendix~\ref{app:lattice} for Fourier transform conventions), and the self-consistency condition can be written in that basis as
\be\label{eq:scgacondition}
\frac{1}{3} = \frac{1}{4N}\sum_{\vec{q} \in {\rm B.Z}}{\rm Tr} \left[\lambda{\mathsf I} + \beta \sum_nJ_n {\mathsf V}^{(n)}(\vec{q})\right]^{-1}
\ee
where the matrices ${\mathsf V}^{(n)}(\vec{q})$ act within the space of the four sublattices. Their explicit forms are provided in Appendix~\ref{app:matrices}. The solutions for $\lambda$ for various choices of further neighbour exchange are shown in Fig.~\ref{fig:scgasolution}. We discuss analytical bounds on the solution following Eq.~(\ref{bounds}),
after the physical significance of $\lambda$ as the stiffness of a Coulomb-phase action has been made clear.

The correlation functions in reciprocal space between different sublattices are
\be\label{eq:sublatticecorrelations}
\langle s_{\mu}(\vec{q})s^*_{\nu}(\vec{q})\rangle = \left[\lambda{\mathsf I} + \beta \sum_nJ_n {\mathsf V}^{(n)}(\vec{q})\right]^{-1}_{\mu\nu}\,
\ee
and the static structure factor is the sum of all 16 such correlators. For the nearest neighbour model this can be obtained analytically at all temperatures in a palatable form: at $T=0$, it is provided by \textcite{IsakovGregor04} (with a different sublattice labeling convention) and for completeness we provide the full, finite temperature form in Eq.~(\ref{eq:nnexactstructure}) in Appendix~\ref{app:other}. In the general case where further neighbour terms are not all zero and analytic expressions are unilluminating, spin correlations at any temperature can be straightforwardly obtained numerically from \refeq{eq:sublatticecorrelations}.

\subsection{Mapping to long wavelength description}\label{sec:sub:longwavelength}
Within the SCGA, Eqns.~(\ref{eq:sublatticecorrelations}) and (\ref{eq:scgacondition}) give a full description of spin correlations in the paramagnetic phase. A more physical understanding comes from studying the long wavelength physics in a way that allows comparisons with related models, including spin ice, interacting dimer models, and more generally any system with a Coulomb phase, such as those discussed in Ref.~\onlinecite{Henley09} and references therein.

In our context, the long wavelength degrees of freedom emerge naturally after interpreting spin configurations as configurations of a lattice vector field defined on the bonds of the diamond lattice. The mapping\cite{IsakovGregor04,Henley05} is constructed as follows.
Each site on a pyrochlore lattice is associated with a diamond bond, whose midpoint it is; the diamond lattice is bipartite so its bonds are orientable in a consistent way across the lattice. Therefore each spin component can be interpreted as the local magnitude of a lattice vector field directed along diamond bonds.
The merit of this description is that the vector field obtained is divergenceless in the limit $\beta J_1\rightarrow\infty$ at fixed $\beta J_{n>1}$.

More precisely, for each spin component, labeled by $\alpha$, define the flux field $\vec{B}^{\alpha}$ by $\vec{B}^{\alpha}(\vec{r}) = S^{\alpha}(\vec{r})\vec{e}(\vec{r})$ where $\vec{e}(\vec{r})$ is a (real space) vector oriented along the diamond bond passing through the pyrochlore lattice site $\vec{r}$.
%
%
The mapping has a straightforward implementation as a change of basis in reciprocal space:
\be\label{eq:basischange}
\begin{pmatrix}
M^{\alpha}(\vec{q})\\
B^{\alpha}_x(\vec{q})\\
B^{\alpha}_y(\vec{q})\\
B^{\alpha}_z(\vec{q})\\
\end{pmatrix}
=\frac{1}{2}
\begin{pmatrix}
1 & 1 & 1 & 1\\
1 & 1 & -1 & -1\\
1 & -1 & 1 & -1\\
1 & -1 & -1 & 1\\
\end{pmatrix}
\begin{pmatrix}
S^{\alpha}_1(\vec{q})\\
S^{\alpha}_2(\vec{q})\\
S^{\alpha}_3(\vec{q})\\
S^{\alpha}_4(\vec{q})\\
\end{pmatrix}\;.
\ee
The directions of the vectors $\vec{e}(\vec{r})$ and the sublattice labeling are shown in Fig.~\ref{fig:lattice}. The vectors $\vec{e}(\vec{r})$ are taken to have length $\sqrt{3/4}$ (in contrast to the convention of unit length by \textcite{IsakovGregor04}) so that the definition of the mapping is consistent with the orthogonal change of basis in \refeq{eq:basischange}. At $\vec{q}=0$, the three spatial components of $\vec{B}$ are the three staggered magnetization order parameters which would describe a transition to $\vec{q}=0$ \Neel\ order as referred to in Refs.~\onlinecite{ChernFennieTchernyshyov06,ChernMoessnerTchernyshyov08}. The experimentally accessible quantity is the structure factor which, we define as $S(\vec{q})=4\langle |M(\vec{q})|^2\rangle$.

We are interested in the small $\vec{q}$ behaviour of the vector field $\vec{B}(\vec{q})$ but the experimental probe is via the linear combination $M(\vec{q})$. Consequently, the long wavelength physics does not appear near $\vec{q}=0$ in the structure factor, but rather near $\vec{q}=\vec{K}$ for special vectors $\vec{K}$, which satisfy $M(\vec{q}+\vec{K}) \propto B_i(\vec{q})$ for some spatial direction $i$. The $\vec{K}$ are a subset of the reciprocal lattice vectors (see Appendix \ref{app:pinchpoints}).

Using this mapping, we now turn to the long wavelength physics of the model defined in Sec.~\ref{sec:sub:scga}. We consider explicitly the case of $J_{3a}=J_{3b} \equiv J_3$ with all other further neighbour interactions zero; it is straightforward to carry out a similar analysis with any other choice of further neighbour terms.

At small $\vec{q}$ and $\beta J_1 \gg 1$, the linear combination $M(\vec{q})$ is nearly frozen out and within the SCGA we can legitimately read off a small $\vec{q}$ theory for the field $\vec{B}$ by restricting the interaction to the $3\times3$ subblock which acts between the components of $\vec{B}$, providing $\lambda$ is still calculated correctly using the full theory, \refeq{eq:scgacondition}. Further details are provided in Appendix \ref{app:matrices}. By expanding the interactions to second order in $\vec{q}$ we obtain as a long wavelength theory
\begin{eqnarray}
\label{eq:smallqhamiltonian}
\beta {\cal H} = \frac{1}{2}\sum_{\vec{q}}
\lambda|\vec{B}(\vec{q})|^2
 &+& \beta J_1a^2|\vec{q}\cdot\vec{B}(\vec{q})|^2\nonumber\\
&-& 8\beta J_3a^2 q^2|\vec{B}(\vec{q})|^2\,.
\end{eqnarray}
The length scale $a$ is the nearest neighbour distance on the pyrochlore lattice. It is clear from \refeq{eq:smallqhamiltonian} that $J_1$ acts purely to enforce the constraint $\vec{q}\cdot\vec{B} = 0$.

We now consider the flux correlations that follow from Eq.~(\ref{eq:smallqhamiltonian}).
The $q=0$ flux fluctuations are straightforwardly obtained as
\be\label{eq:globalfluxfluctuations}
\langle B_i(\vec{q})B^*_j(\vec{q})\rangle_{\vec{q}=\vec{0}} = \frac{1}{\lambda}\delta_{ij}
\ee
from which we identify the parameter $\lambda$ as the dimensionless stiffness of the flux fields in the long wavelength description. Since its original role as a Lagrange multiplier is to imitate the restrictions on phase space due to the fixed spin length, it is natural that it appears also as the stiffness which is entropic in origin. The correlation function in \refeq{eq:globalfluxfluctuations} is directly accessible from simulations and the data points in Fig.~\ref{fig:scgasolution} follow from this relation; the same relation is used to measure the stiffness in simulations of dimer models (see e.g. Eq.~(2) in Ref.~\onlinecite{AletMisguich06}). It is impressive that the SCGA captures behaviour in the Heisenberg model so accurately.

The correlator for longitudinal flux fluctuations is
\be\label{eq:longitudinalfluxcorrelations}
\langle B_x(q_x)B^*_x(q_x)\rangle_{q_y=q_z=0} = \frac{1}{\lambda}\left[\frac{\xi_\parallel^{-2}}{\xi_\parallel^{-2}+q_x^2}\right]
\ee
where we introduce the longitudinal correlation length $\xi_\parallel = a\sqrt{\beta(J_1-8J_3)/\lambda}$ (whose explicit form of course depends on which further neighbour exchanges are included). The inverse correlation length is the width of pinch points in reciprocal space.
In the limit $\beta J_1\rightarrow\infty$ at fixed $\beta J_3$, flux fields are strictly solenoidal and $\xi_\parallel$ diverges. More generally, we emphasise that $\xi_{\parallel}$ depends on $\beta$ both explicitly and through the $\beta$-dependence of $\lambda$, via \refeq{eq:scgacondition}. 

The transverse flux correlations are
\be
\langle B_z(q_x)B^*_z(q_x)\rangle_{q_y=q_z=0} = \frac{1}{\lambda}\left[\frac{1}{1-\frac{8\beta J_3a^2}{\lambda}q_x^2}\right]\;.
\ee
For $J_3$ negative, the length $\xi_\perp = a\sqrt{8\beta |J_3|/\lambda}$ is the transverse correlation length for flux fluctuations. For positive $J_3$, $\xi_\perp$ is not a correlation length: instead it sets a length scale at which  higher order terms in $q$ must be included in the long wavelength theory.

\textcite{YoungbloodAxe81} considered a phenomenological Landau free energy for polarization fluctuations in ferroelectrics, essentially identical to \refeq{eq:smallqhamiltonian}. Our discussion differs in that \refeq{eq:smallqhamiltonian} is derived directly from a microscopic model, the parameter $\lambda$ is allowed to flow, and moreover both positive and negative coupling to transverse gradient terms are considered.

\begin{figure*}
\includegraphics{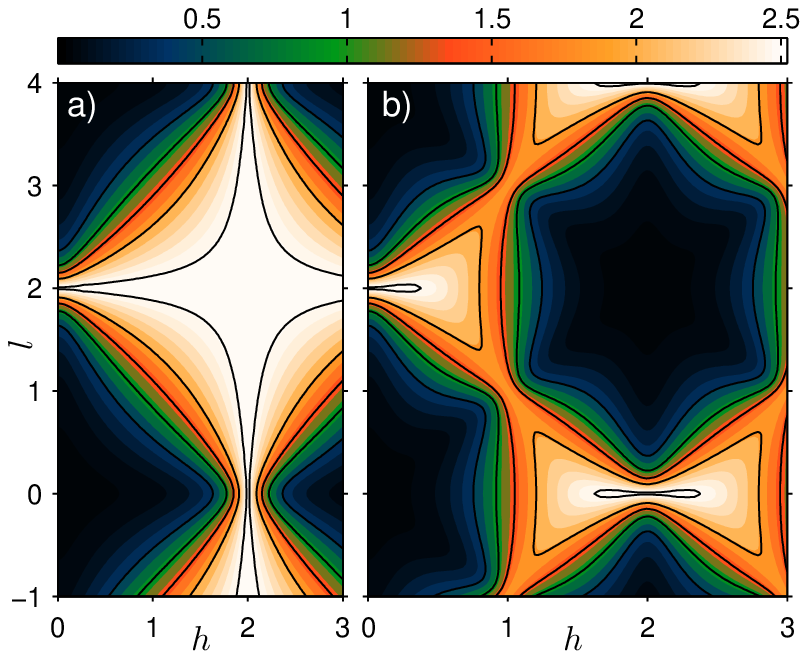}\hfill\includegraphics{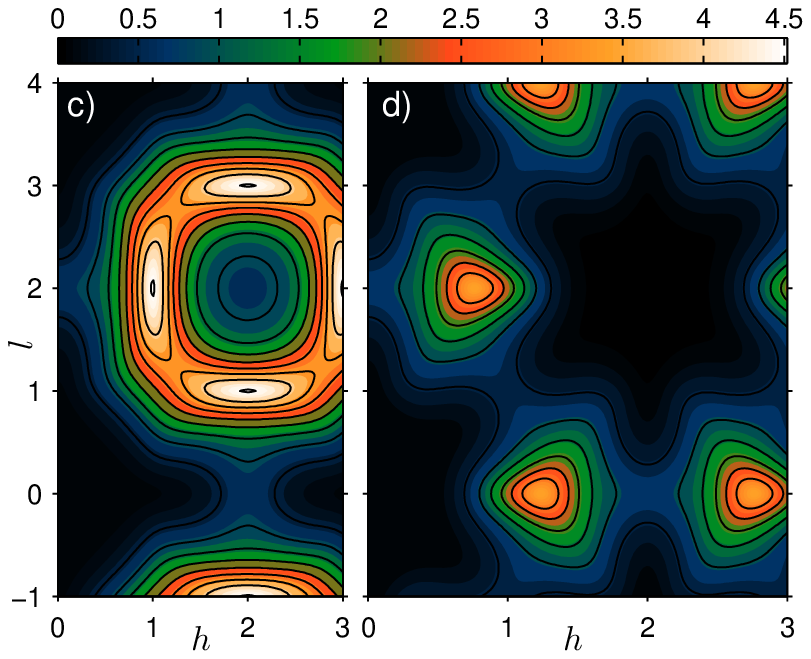}
\caption{\label{fig:afmj3correlations}(color online) Structure factor $S(\vec{q})$ at $T = 0.1J_1$. (a) and (b): NN interactions only; (c) and (d): $J_3=0.025J_1$. (a) and (c): $(h0l)$ plane;  (b) and (d): $(hhl)$ plane. Contour lines are at the levels marked on the color scale. Compared to a nearest neighbour model, the effect of AFM $J_3$ is, first, to suppress and broaden the pinch points, and second, to shift spectral weight so that behaviour resembles more the hexagonal cluster scattering. MC simulations yield plots that are indistinguishable by eye.}
\end{figure*}

Having established that $\lambda$ has a physical interpretation as the Coulomb phase coupling constant, we now turn to bounds on the value of $\lambda$ which solves \refeq{eq:scgacondition}. Let $\varepsilon_{\rm min}$ be the minimum eigenvalue of $\sum_nJ_n {\mathsf V}^{(n)}(\vec{q})$ over all $\vec{q}$. The requirement that the quadratic form in \refeq{eq:scgahamiltonian} is positive-definite demands that $\lambda + \beta\varepsilon_{\rm min} > 0$. Consequently we have
\be\label{bounds}
\lambda > -\beta\varepsilon_{\rm min}\;.
\ee
Taking an AFM $J_1$, we consider two special cases. First,
for an AFM $J_3$ and all other $J_{n>1}=0$, we have $\varepsilon_{\rm min}=-16J_3$. Second, for a FM $J_2$ and all other $J_{n>1}=0$, the position of the minimum in reciprocal space is given by \textcite{ChernMoessnerTchernyshyov08} in their Eq.~(4), with $\varepsilon_{\rm min}\simeq-8.57|J_2|$ for $|J_2|\ll J_1$.


\section{Consequences of AFM $J_3$ or FM $J_2$}\label{sec:fmj2afmj3}
\subsection{Paramagnetic phase}
Within the paramagnetic phase, AFM $J_3$ or FM $J_2$ act to suppress pinch points, and to promote cluster-like correlations. These effects are shown in Fig.~\ref{fig:afmj3correlations}, which should be compared, for example, with Fig. 3 of Ref.~\onlinecite{LeeBroholm02}. The behaviour illustrated here is one of our main results. Its origin is discussed in this section.
\subsubsection{Consequences for pinch point visibility}
\begin{figure}
\includegraphics{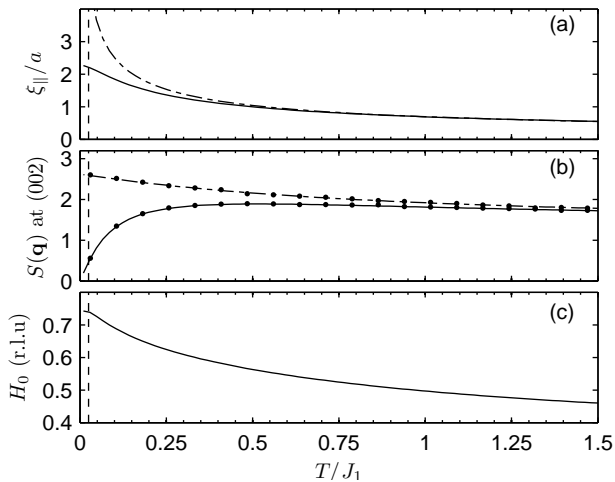}
\caption{\label{fig:fmj2mixed} Temperature dependence of quantities characterising the scattering, for $J_2=-0.025J_1$. (a) Coulomb phase correlation length $\xi$ in units of site spacing for NN model (broken line) and for model with non-zero $J_2$ (solid line). (b) Amplitude of correlation function at pinch-point, for both models. (c) Position $H_0$ of maximum in scattering, as described in text. Dashed line indicates temperature $T=J_2$: the effect of non-zero $J_2$ extends to $T\gg J_2$. Curves are from SCGA. Points are from simulations on 2048 spins.}
\end{figure}

The pinch points are the signature of algebraic correlations in a Coulomb phase. There are three ways in which they may become less visible. (i) Further neighbour interactions or finite temperature endow the pinch points with a width set by the inverse of the longitudinal correlation length. (ii) The absolute amplitude of the pinch points is set by the inverse of the flux field stiffness. (iii) The visibility of the pinch points depends also on the relative weight at the pinch points compared to that at the maximum in scattering. We address these mechanisms quantitatively below.
(i) For AFM $J_3$, the longitudinal correlation length is $a\sqrt{\beta(J_1-8J_3)/\lambda}$. The lower bound on $\lambda$ puts an upper bound on the correlation length of $a\sqrt{(J_1/16J_3-1/2)}$. Even with $J_3\simeq J_1/150$, the correlation length never exceeds 3 site spacings and the pinch points are thus never sharp. For FM $J_2$, a similar argument gives an upper bound on the correlation length of approximately $a\sqrt{(2+J_1/|J_2|)/8.6}$. The effect on the correlation length is shown in Fig.~\ref{fig:fmj2mixed}(a) for $J_2=-0.025J_1$.
(ii) Since the absolute amplitude of the pinch points is set by $1/\lambda$, the lower bound on $\lambda$ puts an upper bound on the absolute intensity of pinch point correlations. For AFM $J_3$, recall that $\lambda > 16\beta J_3$, whereas for FM $J_2$, we have $\lambda > 8.6\beta J_2$, for $|J_2|\ll J_1$. The pinch point intensity is shown in Fig.~\ref{fig:fmj2mixed}(b) for $J_2 = -0.025 J_1$.
(iii) Moreover, as spectral weight shifts away from the pinch points to other wavevectors, the relative amplitude given by the ratio of pinch point correlations to the maximum in the correlation function decreases even more rapidly.
These three mechanisms indicate how pinch points may both be suppressed in amplitude and broadened in width by further neighbour interactions.

\subsubsection{Hexagonal cluster-like correlations}
We have seen how further neighbour terms suppress spectral weight at the pinch points. We turn now to why antiferromagnetically correlated loops of spins such as hexagonal clusters may appear in a model with further neighbour terms. The mechanism is straightforward to explain in the long wavelength description. For simplicity, we consider first the limit $\beta J_1\rightarrow\infty$ (at finite $\beta J_2$). In this case, pinch points remain infinitely sharp since the local constraint of zero divergence is enforced rigorously, but their amplitude is still suppressed since the flow of the stiffness with temperature is unaffected. The further neighbour terms couple to transverse gradients of the flux fields and the effect of AFM $J_3$ or FM $J_2$ is to promote transverse gradients in the flux field. In the continuum picture, maximising transverse gradients requires having as many small loops of flux as possible; when understood in terms of the lattice geometry, the smallest loops of flux available are the hexagonal clusters. Considered microscopically, it is also clear from Fig.~\ref{fig:lattice} that an AFM $J_{3b}$ is likely to induce antiferromagnetically correlated hexagons of spins.

\begin{figure}
\includegraphics{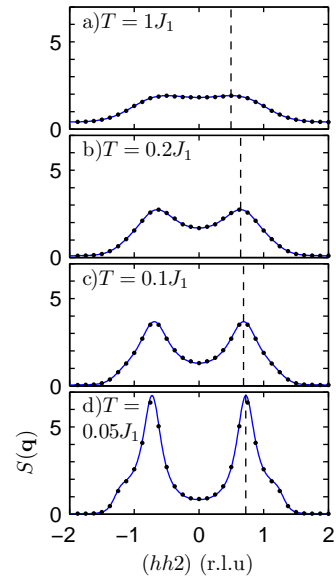}
\caption{\label{fig:fmj2cuts} Structure factor along $(hh2)$ for $J_2=-0.025J_1$. Lines are from SCGA, points from simulations on 16384 spins. The position $H_0$ at each temperature is indicated by dashed line.}
\end{figure}

Following Ref.~\onlinecite{KamazawaParkLee04}, we define $H_0$ as the temperature-dependent position of the maximum in scattering along the (hh2) direction (see dashed lines in Fig.~\ref{fig:fmj2cuts}). For FM $J_2$, at T=0, $H_0$ coincides with $h^*$ given by \textcite{ChernMoessnerTchernyshyov08} in their Eq.~(4), the position of the minimum eigenvalue of the exchange matrix. The evolution of $H_0$ with temperature extracted from the SCGA is shown in Fig.~\ref{fig:fmj2mixed}(c). Qualitatively similar behaviour for $H_0$ is found experimentally in \CdFeO\ (see Fig. 4 in Ref.~\onlinecite{KamazawaParkLee04}). In contrast, in the NN model (at least classically) the scattering maxima are always found at the pinch points, i.e $H_0=0$, so the NN model does not provide an explanation for why the maximum in scattering is elsewhere.

\begin{figure}
\includegraphics{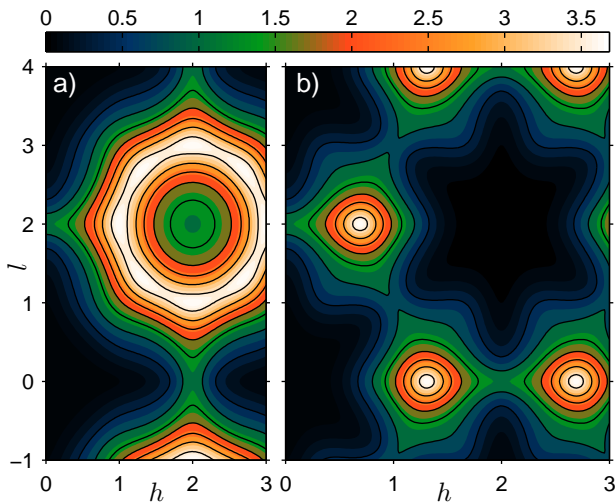}
\caption{\label{fig:fmj2correlations}(color online) Structure factor at $T=0.1J_1$ with $J_2=-0.025J_1$. Panels as in Fig.~\ref{fig:afmj3correlations}.}
\end{figure}

In Fig.~\ref{fig:fmj2correlations}, we show the structure factor on the same planes of reciprocal space as in Fig.~\ref{fig:afmj3correlations}, but for the case of a FM $J_2$ rather than an AFM $J_3$. The qualitative effect is similar, but the characteristic scattering shapes look less like the hexagonal form factor used to explain the scattering in many chromites.

\subsection{Ordering transitions}
For $J_3\ll J_1$, we observe a first order phase transition at $T_c\simeq2.2J_3$ in our MC simulations (eg. see upper inset in Fig.~\ref{fig:scgasolution}). We have not attempted to characterize the ordered state since our focus is on the strongly correlated paramagnet in the window $T_c<T<J_1$. A phase transition is not predicted by mean field calculations\cite{ReimersBerlinskyShi91}, nor within the theory of Sec.~\ref{sec:sub:scga} since the interaction minima has a 1-dimensional degeneracy in reciprocal space.

In the phase diagram presented by \textcite{ChernMoessnerTchernyshyov08}, ferromagnetic $J_2$ leads to incommensurate order at low temperatures consistent with mean field calculations\cite{ReimersBerlinskyShi91}, but also an intermediate collinear regime, stabilised by thermal fluctuations, with first order transitions out of the paramagnetic phase. The transition temperature is approximately at $T_c\simeq J_2$, but see Ref.~\onlinecite{ChernMoessnerTchernyshyov08} for details.

\section{Consequences of AFM $J_2$ or FM $J_3$}\label{sec:afmj2fmj3}
Although the main focus of this paper is further neighbour terms with their sign chosen to suppress pinch point scattering, we now also consider further neighbour terms with the opposite sign. These enhance pinch point scattering.
\subsection{Paramagnetic phase}
It is clear from \refeq{eq:smallqhamiltonian} that in the flux description an antiferromagnetic $J_2$ or FM $J_3$ penalizes transverse gradients of the flux fields. Qualitatively then, one expects spectral weight to accumulate at small $q$ as short wavelength fluctuations are suppressed, developing into Bragg peaks as the transverse flux correlation length diverges at a phase transition.

\begin{figure}
\includegraphics{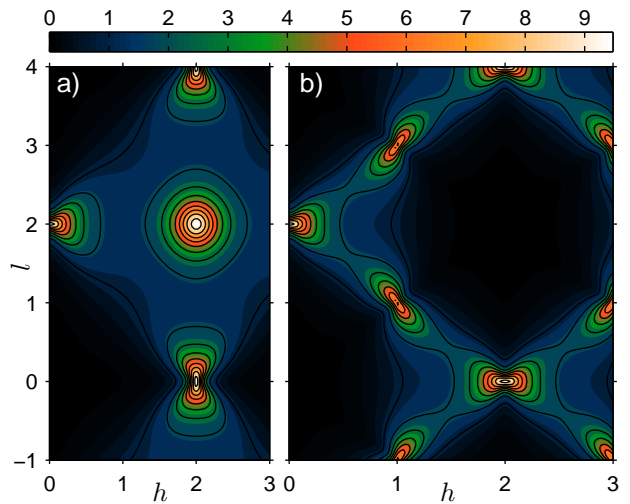}
\caption{\label{fig:afmj2correlations} (color online) Structure factor $S(\vec{q})$ at $T = 0.1J_1$ and $J_2 = 0.025 J_1$ in (a) the $(h0l)$ plane and (b) the $(hhl)$ plane. Contour lines are at the levels given in the color scale. Positive $J_2$ increases pinch point scattering, and eventually leads to \Neel~order. The shapes around the pinch points are almost identical to those discussed by \textcite{YoungbloodAxe81}.}
\end{figure}

As seen from the numerical solution of \refeq{eq:scgacondition} shown in Fig.~\ref{fig:scgasolution}, FM $J_3$ causes the stiffness $\lambda$ to decrease with cooling, and the same behaviour is found with AFM $J_2$. The strength of flux fluctuations is given by the inverse stiffness as in Eq.~(\ref{eq:globalfluxfluctuations}) and consequently this increases. In Fig.~\ref{fig:afmj2correlations} we show spin correlations in the paramagnetic phase at the same value of $\beta J_1$ as in Fig.~\ref{fig:fmj2correlations}: with antiferromagnetic $J_2$, pinch points are intensified and sharper than for the nearest neighbour model as expected.
\textcite{YoungbloodAxe81} considered the Maxwell action with terms penalising transverse flux gradients and obtained a similar effect.

\subsection{Ordering transitions}
With an AFM $J_2\ll J_1$, the full model, \refeq{eq:fullhamiltonian}, has a first order transition to a $\vec{q}=0$ \Neel-ordered phase at $T\simeq3.2J_2$\cite{ChernMoessnerTchernyshyov08}. Within the framework of Sec.~\ref{sec:sub:scga}, the transition is second order and occurs at a lower temperature ($T_c\simeq 2.44 J_2$ for $J_1/J_2=100$): the one at which the minimum eigenvalue of the quadratic form in \refeq{eq:scgahamiltonian} vanishes. For FM $J_3\ll J_1$, our simulations reveal a first order transition at $T\simeq 8|J_3|$ (see lower inset of Fig.~\ref{fig:scgasolution}). In both cases, the transition is to a fully flux-polarised state. Similar transitions occur in other Coulomb phase systems: for pyrochlore Heisenberg spins, this is $\vec{q}=0$ \Neel\ order, for interacting dimer models, a transition to a uniform tilted phase, and in the context of polarization fluctuations in paraelectrics\cite{YoungbloodAxe81}, a ferroelectric transition out of the paraelectric phase.

The transverse flux correlation length $\xi_\perp$, which is zero for a pure nearest neighbour model, diverges as $\lambda\rightarrow 0$. Flux fluctuations are correlated over larger regions and the divergence of $\xi_\perp$ marks the transition into an ordered phase.
\section{General combinations of further neighbour interaction}\label{sec:generalcouplings}
\begin{figure}
\includegraphics{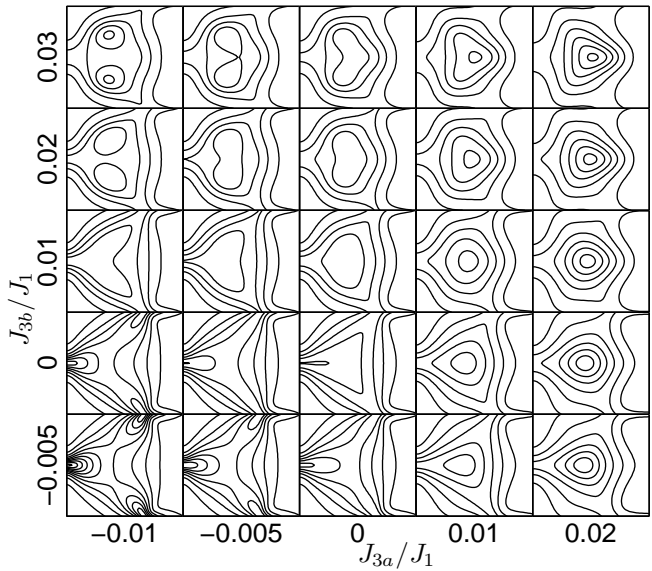}
\caption{\label{fig:j3aj3bcomp} $S(\vec{q})$ in the $(hhl)$ plane for $h\in[0,3/2]$, $l\in[1,3]$ at $T = 0.1J_1$, $J_2=0$ and with $J_{3a}$, $J_{3b}$ interactions treated independently, showing the variety of short range correlations, and the effects on the pinch point at $(002)$.}
\end{figure}

In Secs.~\ref{sec:fmj2afmj3} and \ref{sec:afmj2fmj3} we took both third neighbour couplings as equal, but in general they need not be and here we consider $J_{3a}$ and $J_{3b}$ as distinct. Within the long wavelength description, nothing qualitatively changes since further neighbour terms can only act to add gradient terms while renormalising the stiffness. However, the detailed nature of the short range paramagnetic correlations is not generic and depends sensitively on the ratios of the different further neighbour couplings. Since this balancing act is separate for each candidate material which our ideas hope to describe, we provide a survey for several different combinations. In Fig.~\ref{fig:j3aj3bcomp}, we have plotted paramagnetic correlations in the reciprocal space patch described in the caption for a variety of different $J_{3a}$ and $J_{3b}$ at $T=0.1J_1$. By the arguments of Ref.~\onlinecite{ChernMoessnerTchernyshyov08}, $J_2$ acts as $-J_{3a}$ for large enough $J_1$, so the horizontal axis can be considered approximately to probe $J_{3a}-J_2$.

\section{Dynamics and inelastic scattering}\label{sec:dynamics}
\begin{figure}
\includegraphics{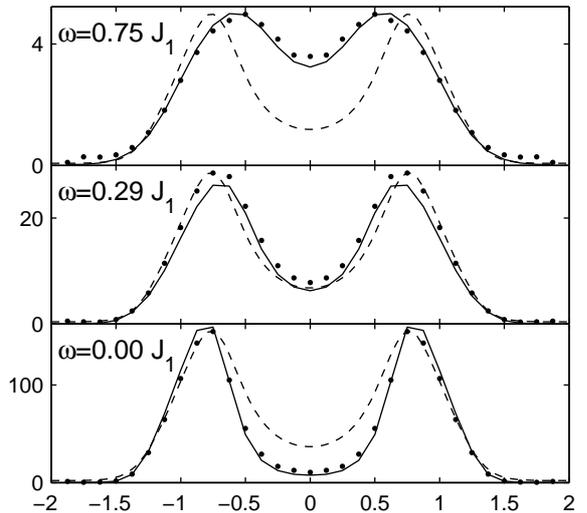}
\caption{\label{fig:j3dynamics} Points: $S(\vec{q},\omega)$ along $(hh2)$ at $T = 0.1J_1$, $J_3=0.025J_1$ from simulations. Solid line: $S(\vec{q},\omega)$ from stochastic model of Ref.~\onlinecite{ConlonChalker09prl} with further neighbour terms. Broken line: $S(\vec{q})$ rescaled for comparison. Pinch points are more suppressed at small-$\omega$, and conversely less suppressed at high-$\omega$.}
\end{figure}

We have so far considered only equal time correlations but further neighbour (FN) terms will also affect dynamics. There is an intrinsic precessional dynamics associated with the Heisenberg spins in \refeq{eq:fullhamiltonian}. Nevertheless, we have shown elsewhere\cite{ConlonChalker09prl} for the NN model that the low temperature dynamics is dominated by relaxational modes and is very well captured by a purely relaxational stochastic model based on the SCGA of Sec.~\ref{sec:sub:scga}. We find this statement about the low temperature paramagnet remains true even with FN terms. As evidence for this, we present in  Fig.~\ref{fig:j3dynamics} the results of molecular dynamics simulations using the same methods as in Ref.~\onlinecite{ConlonChalker09prl} for the dynamic structure factor, plotted at 3 different frequencies $\omega$. On the same plot, we show the prediction obtained from including FN terms in the stochastic model. Compared to the static result, the pinch point scattering at $(002)$ is more suppressed at small $\omega$, and conversely, less suppressed at higher $\omega$. The implications for quasi-elastic scattering experiments at $\omega\simeq0$ are that we expect the effects of FN terms to be more conspicuous (and conversely at larger $\omega$, to become less conspicuous) when compared to the static ($\omega$-integrated) scattering which is the main focus of this paper.

The enhanced suppression can be understood by considering the stochastic model in Ref.~\onlinecite{ConlonChalker09prl}. Within that model the dynamic structure factor is understood as the sum of contributions from the four bands of the interaction matrix, given by
\be\label{eq:dynamicstruc}
S(\vec{q},\omega) = \sum_{\mu = 1}^{4}g_{\mu}(\vec{q})f_{\mu}(\vec{q},\omega)\,,
\ee
where the $g_{\mu}(\vec{q})$ are form factors, independent of temperature, and where the thermal weight and dynamics for each band are contained in the functions $f_{\mu}(\vec{q},\omega)$. The quadratic form in \refeq{eq:scgahamiltonian} has eigenvalues $\alpha_{\mu}(\vec{q})$, labeled by band index $\mu$, which determine relaxation rates $\Gamma_{\mu}(\vec{q})\propto\alpha_{\mu}(\vec{q})$
and these enter the dynamic structure factor through the relation
\be\label{eq:dynamicform}
f_{\mu}(\vec{q},\omega) = \alpha_{\mu}^{-1}(\vec{q})\frac{2\Gamma_{\mu}(\vec{q})}{\omega^2+\Gamma_{\mu}(\vec{q})^2}\,.
\ee
By integrating over $\omega$ we obtain the static structure factor,
which has the form of \refeq{eq:dynamicstruc}, but with $\alpha_{\mu}^{-1}(\vec{q})$ in place of
$f_{\mu}(\vec{q},\omega)$.

With only nearest neighbour interactions, the two lowest bands (which are dominant at low temperature) are dispersionless and degenerate, which has two consequences: the $\vec{q}$-dependence arises entirely from the form factors in \refeq{eq:dynamicstruc} since the relevant $f_\mu$ are independent of $\vec{q}$, and the $\omega$-dependence approximately factorises out. With FN interactions, the lowest bands acquire dispersion and the relevant $f_\mu$ are no longer independent of $\vec{q}$. If their ratio in the static case for band $\mu$ at $\vec{q}_1$ and $\vec{q}_2$ is
$r$, then in the dynamic case with $\omega=0$ it is $r^2$. This is why the suppression effects we have discussed in this paper, which correspond to $r\neq1$, are enhanced in quasi-elastic scattering. Further discussion of dynamical correlations is given in Ref.~\onlinecite{ConlonThesis}.

\section{Estimates of further neighbour exchange}\label{sec:exchangeestimates}
Evidence from {\it ab initio} calculations suggests third neighbour couplings are important in a number of relevant compounds. In a study of $A{\rm Cr_2}X_4$ spinels including zinc and cadmium chromite, \textcite{Yaresko08} finds that AFM coupling between third neighbours is important in all the compounds considered, and that the ratio $J_{3b}/J_{3a}\simeq 0.5$ for Zn and Cd. (Our $J_{3a}$ and $J_{3b}$ are labeled as $J_{3'}$ and $J_{3''}$ in Ref.~\onlinecite{Yaresko08}). A separate {\it ab initio} study of \CdCrO~finds $J_3/J_1\simeq 0.3$ where both third neighbour couplings are taken with the same strength\cite{ChernFennieTchernyshyov06}. In both \ZnFeO\ and \CdFeO, \textcite{Cheng08} also reports that third neighbour interactions are estimated to be much stronger than second-neighbour interactions and on the same order in magnitude as first-neighbour interactions. This result is in broad agreement with a study of third neighbour interactions in \ZnFeO\ (Ref.~\onlinecite{YamadaKamazawa02}).

Experimental estimates of nearest neighbour and next nearest neighbour exchange from susceptibility measurements\cite{KantDeisenhoferRudolf09} find for \CdCrO\ a FM $J_2 \simeq -0.3 J_1$ but for \ZnCrO\ an AFM $J_2\simeq0.15J_1$. From our results in Sec.\ref{sec:afmj2fmj3}, the latter would imply scattering quite different from that observed; we have not examined the reliability of the model used in Ref.~\onlinecite{KantDeisenhoferRudolf09} in the presence of third neighbour terms, but believe that \ZnCrO\ is unlikely to be described by a Heisenberg model with AFM $J_2$ without other further neighbour terms included.

\section{Discussion}\label{sec:discussion}
The appearance of antiferromagnetically correlated hexagonal clusters in the spinels is certainly puzzling in the context of a pure nearest neighbour model, since properties of the NN model have been reliably calculated and give different behaviour. In Sec.~\ref{sec:exchangeestimates}, we discussed evidence for further neighbour exchange and found that AFM third neighbour exchange is important in many of the compounds we have been discussing. We should also address other possibilities.

It has been suggested that spin clusters are induced by magnetoelastic coupling\cite{TomiyasuSuzuki08}. However, we note that among the compounds where spin clusters have been reported, \CdCrO\ undergoes a $c$-axis elongation\cite{ChungMatsuda05}, whereas both \ZnCrO\ and \MgCrO\ undergo a $c$-axis contraction, and \CdFeO\ has no structural transition at all; it is not clear how magnetoelastic coupling which leads to different structural transitions should explain the cluster-like scattering common to all of these materials. Moreover, scattering in $\rm Y(Sc)Mn_2$ is broadly similar to the spinels\cite{BallouLelievre96}, but we expect the nature of spin lattice coupling to be different, and this throws further doubt over the idea that magnetoelastic coupling may be universally responsible for cluster-like scattering.

We should also discuss the temperature dependence of our results. The images in Figs.~\ref{fig:afmj3correlations} and \ref{fig:fmj2correlations} are for a single temperature, where the hexagonal cluster-like scattering is most visible; at lower temperatures however, the further neighbour terms induce magnetic phase transitions as discussed in the text. If indeed further neighbour terms are required to explain the hexagonal cluster scattering, we may also expect to observe the associated magnetic phase transitions. In the chromites, we can avoid the issue by assuming the magnetic phase transition is preempted by the structural transitions driven by the independent mechanism of spin-lattice coupling; this is not entirely satisfactory as an explanation, since \CdFeO\ shows neither a structural transition, nor development of any Bragg peaks at any temperature. It is possible that quantum fluctuations or disorder can stabilise the paramagnetic phase\cite{SaundersChalker08}, but it is hard to be conclusive about a range of materials at once, and the role of further neighbour exchange should be assessed individually for different materials.

We have presented further neighbour exchange as the simplest extra interaction which gives hexagonal cluster scattering, but further neighbour exchange is not the only perturbation one could imagine. Others include Dzyaloshinsky-Moriya interactions \cite{KotovElhajal05,ElhajalCanalsSunyer05} and biquadratic exchange, which can appear microscopically or as an effective term taking into account quantum fluctuations\cite{LarsonHenley08}. In future work it would be interesting to consider the effects of these other perturbations on the paramagnetic phase. The present results are sufficient to show that further neighbour exchange of a realistic strength can dramatically alter the diffuse scattering and should not be ignored in comparing experimental results to simple theoretical models.

Clusters should be understood, then, as short range order in a strongly correlated state. A similar view is expressed by \textcite{Yavorskii08} when considering cluster-like scattering in spin ices; their analysis including further neighbour terms suggests that cluster-like scattering is the property of a strongly correlated liquid state which is sensitive to weak perturbations rather than due to the emergence of `real' clusters.

Our results potentially explain why a clear signature of algebraic correlations is missing in candidates for pyrochlore Heisenberg magnets. Small further neighbour interactions are sufficient to wash out the pinch points. Our results may have implications for other pyrochlore models, but in Ising models where the ice-rules are exponentially enforced, we do not expect the dramatic broadening we observe in the Heisenberg model. Nevertheless, the suppression of pinch point scattering amplitude is expected, and we note that in Ref.~\onlinecite{FennellDeenWildes09} such a suppression is reported in the spin ice \HoTiO.

In summary, we have studied the effect of further neighbour interactions on the low temperature paramagnetic phase of the frustrated Heisenberg antiferromagnet on the pyrochlore lattice. Further neighbour terms induce transitions to ordered phases, but they also have a striking effect on paramagnetic correlations. In the description of the low temperature paramagnet as a Coulomb phase, further neighbour terms cause the Coulomb phase coupling constant, or stiffness, to flow. Pinch points in diffuse scattering have their amplitude and width controlled by the stiffness, and even very weak further neighbour terms can cause pinch points to be suppressed in amplitude and broadened. With FM $J_2$ or AFM $J_3$, paramagnetic scattering is altered to resemble more the hexagonal cluster scattering that is often observed in experimental systems, notably spinels. Further neighbour terms then provide a mechanism for the previously unexplained cluster-like scattering in frustrated spinels.

\begin{acknowledgments}PC acknowledges helpful discussions with M.~Gingras and S.~Bramwell.
This work was supported in part by EPSRC Grant Number EP/D050952/1.
\end{acknowledgments}

\appendix
\section{Lattice definitions}\label{app:lattice}
We take throughout a conventional cubic unit cell of side $1$, and use $a$ to refer to the pyrochlore nearest neighbour distance, so that $a^2=1/8$. The midpoints of the tetrahedra form a diamond lattice with site spacing $\sqrt{3/16}$.

The pyrochlore lattice is a face centered cubic (fcc) lattice decorated with a tetrahedron at each fcc site. Using the following choice of fcc basis vectors
\begin{equation}
\vec{a}_1 = \frac{1}{2}(\vec{e}_y+\vec{e}_z)\;,\quad\vec{a}_2 = \frac{1}{2}(\vec{e}_z+\vec{e}_x)\;,\quad
\vec{a}_3 = \frac{1}{2}(\vec{e}_x+\vec{e}_y)\;,
\end{equation}
then (up to an arbitrary displacement) the corners of the tetrahedra appear at positions
\begin{equation}
\vec{c}_{\mu} \in \left\{ \vec{0}\;,\frac{\vec{a}_1}{2}\;,\frac{\vec{a}_2}{2}\;,\frac{\vec{a}_3}{2} \right\}
\end{equation}
where $\mu$ labels the four sublattices and runs from $1-4$, as in Fig.~\ref{fig:lattice}.
The Fourier transforms we use are
\begin{align}\label{eq:fouriertransform}
S_{\mu}(\vec{r}) &= \frac{1}{\sqrt{N}}\sum_{\vec{q}\in\text{B.Z}}S_{\mu}(\vec{q})e^{i\vec{q}.(\vec{r}+\vec{c}_{\mu})}
\notag\\
S_{\mu}(\vec{q}) &= \frac{1}{\sqrt{N}}\sum_{\vec{r}}S_{\mu}(\vec{r})e^{-i\vec{q}.(\vec{r}+\vec{c}_{\mu})}
\end{align}
where $\vec{r}$ runs over $N$ fcc lattice sites so that there are $4N$ spins in total.

\section{Interaction matrices}\label{app:matrices}
As described in the main text, with $A^{(n)}_{ij}$ denoting the adjacency matrix that is 1 if sites $i$ and $j$ are $n$th neighbours and zero otherwise, the interaction matrices are chosen to be $V^{(1)}_{ij} = A^{(1)}_{ij}+2\delta_{ij}$; $V^{(2)}_{ij} = A^{(2)}_{ij}+4\delta_{ij}$; $V^{(3a)}_{ij} = A^{(3a)}_{ij}-6\delta_{ij}$ and $V^{(3b)}_{ij} = A^{(3b)}_{ij}-6\delta_{ij}$. This choice of diagonal entries is made so that when expressed in reciprocal space according to \refeq{eq:fouriertransform} and using the basis change in \refeq{eq:basischange}, the $3\times3$ subblock which acts among the flux components vanishes at $\vec{q}=0$. This ensures that $\lambda$ can be interpreted as the stiffness of the flux fluctuations.

The matrix elements for the interactions can be found in Ref.~\onlinecite{ReimersBerlinskyShi91}, although in a different basis so we provide them here using our conventions.
We first define the fcc displacement vectors $\vec{t}_{\mu\nu} = 2(\vec{c}_{\mu}-\vec{c}_{\nu})$ where the $\vec{c}_\mu$ are as in Appendix.~\ref{app:lattice}.
After Fourier transform the matrix elements of the adjacency matrices are given by
\begin{equation*}
A^{(1)}(\vec{q})_{12} = 2 \cos\left(\frac{\vec{q}\cdot\vec{t}_{12}}{2}\right)
\end{equation*}
and
\begin{equation*}
A^{(2)}(\vec{q})_{12} = 2 \cos\left(\frac{\vec{q}\cdot(\vec{t}_{13}+\vec{t}_{23})}{2}\right)+2 \cos\left(\frac{\vec{q}\cdot(\vec{t}_{14}+\vec{t}_{24})}{2}\right)
\end{equation*}
with diagonal elements zero, and the rest by permutation.

Furthermore, third neighbour couplings are diagonal giving:
\begin{equation*}
A^{(3a)}(\vec{q})_{11} = 2 \cos\left(\vec{q}\cdot\vec{t}_{12}\right)+2 \cos\left(\vec{q}\cdot\vec{t}_{13}\right)+2 \cos\left(\vec{q}\cdot\vec{t}_{14}\right)
\end{equation*}
and
\begin{equation*}
A^{(3b)}(\vec{q})_{11} = 2 \cos\left(\vec{q}\cdot\vec{t}_{23}\right)+2 \cos\left(\vec{q}\cdot\vec{t}_{24}\right)+2 \cos\left(\vec{q}\cdot\vec{t}_{34}\right)\;.
\end{equation*}

To derive the long wavelength theory in the flux sector, we implement a change of basis as described in the main text using the orthogonal matrix
\begin{equation*}
{\mathsf P}=\frac{1}{2}
\begin{pmatrix}
1 & 1 & 1 & 1\\
1 & 1 & -1 & -1\\
1 & -1 & 1 & -1\\
1 & -1 & -1 & 1\\
\end{pmatrix}
\end{equation*}
and then take the small $\vec{q}$ limit. This procedure gives the following interactions
\begin{widetext}
\begin{equation*}
\mathsf{PV^{(1)}P^T} \simeq \begin{pmatrix}
8-a^2q^2 & -a^2q_yq_z & -a^2q_xq_z & -a^2q_xq_y \\
-a^2q_yq_z & a^2q_x^2 & a^2q_xq_y & a^2q_xq_z \\
-a^2q_xq_z & a^2q_xq_y & a^2q_y^2 & a^2q_yq_z \\
-a^2q_xq_y & a^2q_xq_z & a^2q_yq_z & a^2q_z^2 \\
\end{pmatrix}
\end{equation*}
and
\begin{equation*}
\mathsf{PV^{(2)}P^T} \simeq \begin{pmatrix}
16-6a^2q^2 & 2a^2q_yq_z & 2a^2q_xq_z & 2a^2q_xq_y \\
2a^2q_yq_z & -2a^2q_x^2 & -2a^2q_xq_y & -2a^2q_xq_z \\
2a^2q_xq_z & -2a^2q_xq_y & -2a^2q_y^2 & -2a^2q_yq_z \\
2a^2q_xq_y & -2a^2q_xq_z & -2a^2q_yq_z & -2a^2q_z^2 \\
\end{pmatrix}
+4a^2\begin{pmatrix}
0 & 0 & 0 & 0 \\
0 & q_y^2+q_z^2 & 0 & 0 \\
0 & 0 & q_x^2+q_z^2 & 0 \\
0 & 0 & 0 & q_x^2+q_y^2 \\
\end{pmatrix}
\end{equation*}
and
\begin{equation*}
\mathsf{PV^{(3a)}P^T} \simeq 4a^2\begin{pmatrix}
-q^2 & -q_yq_z & -q_xq_z & -q_xq_y \\
-q_yq_z & -q_x^2 & -q_xq_y & -q_xq_z \\
-q_xq_z & -q_xq_y & -q_y^2 & -q_yq_z \\
-q_xq_y & -q_xq_z & -q_yq_z & -q_z^2 \\
\end{pmatrix}
-4a^2\begin{pmatrix}
0 & 0 & 0 & 0 \\
0 & q_y^2+q_z^2 & 0 & 0 \\
0 & 0 & q_x^2+q_z^2 & 0 \\
0 & 0 & 0 & q_x^2+q_y^2 \\
\end{pmatrix}
\end{equation*}
\begin{equation*}
\mathsf{PV^{(3b)}P^T} \simeq 4a^2\begin{pmatrix}
-q^2 & q_yq_z & q_xq_z & q_xq_y \\
q_yq_z & q_x^2 & q_xq_y & q_xq_z \\
q_xq_z & q_xq_y & q_y^2 & q_yq_z \\
q_xq_y & q_xq_z & q_yq_z & q_z^2 \\
\end{pmatrix}
-4a^2\begin{pmatrix}
0 & 0 & 0 & 0 \\
0 & 2q_x^2+q_y^2+q_z^2 & 0 & 0 \\
0 & 0 & q_x^2+2q_y^2+q_z^2 & 0 \\
0 & 0 & 0 & q_x^2+q_y^2+2q_z^2 \\
\end{pmatrix}
\end{equation*}
\end{widetext}
In each case we have split the interaction into a part which acts like $J_1$ within the flux field subspace, and another part. \textcite{ChernMoessnerTchernyshyov08} argue that $J_2$ and $-J_{3a}$ have the same effect if $J_1\rightarrow\infty$. This fact is evident in the small $q$ expansions above.
If the two symmetry inequivalent third neighbour couplings are taken to have the same strength, as done by \textcite{ReimersBerlinskyShi91}, then the rotated third neighbour interaction matrix is particularly simple:
\begin{equation}
\mathsf{P[V^{(3a)}+V^{(3b)}]P^T} \simeq
-8a^2 {\rm diag}(q^2)\;.
\end{equation}
Integrating out $M(\vec{q})$ in the long wavelength limit leads to couplings at $\mathcal{O}(q^4)$ between the components $B_i(\vec{q})$, which we have ignored in the analysis at small $\vec{q}$ in Sec.~\ref{sec:sub:longwavelength}.
\section{Locations of pinch points in structure factor}\label{app:pinchpoints}
The experimental probe in scattering experiments is the structure factor $S(\vec{q})=4\langle |M(\vec{q})|^2\rangle$. However the long-wavelength physics is in the field $\vec{B}$ near $\vec{q}=0$. In this appendix we determine the places in reciprocal space where signatures of the behaviour of $\vec{B}$ are visible.

For  $\vec{K}$ a reciprocal lattice vector we have
\be
S_\mu(\vec{q}+\vec{K}) = e^{i\vec{K}.\vec{c}_{\mu}}S_\mu(\vec{q})\;.
\ee
Taking ${\vec{b}_i}$ as the reciprocal basis to ${\vec{a}_j}$, then a reciprocal lattice vector
$\vec{K} = n_1\vec{b}_1+n_2\vec{b}_1+n_3\vec{b}_1$ is
\begin{multline}
\vec{K}=2\pi[(n_2+n_3-n_1)\hat{\vec{x}}\\
+(n_3+n_1-n_2)\hat{\vec{y}}+(n_1+n_2-n_3)\hat{\vec{z}}]
\end{multline}
Bearing in mind the definitions of $\vec{c}_\mu$, the phase factors for the different sublattices are ${1,(-1)^{n_1},(-1)^{n_2},(-1)^{n_3}}$. It follows that
\be
M(\vec{q}+\vec{K}) = \frac{1}{4}[v_0M(\vec{q}) + \vec{v}\cdot\vec{B}(\vec{q})]
\ee
where
\be
v_0 = 1+(-1)^{n_1}+(-1)^{n_2}+(-1)^{n_3}
\ee
and
\be
\vec{v} =
\begin{pmatrix}1+(-1)^{n_1}-(-1)^{n_2}-(-1)^{n_3}\\
1-(-1)^{n_1}+(-1)^{n_2}-(-1)^{n_3}\\
1-(-1)^{n_1}-(-1)^{n_2}+(-1)^{n_3}\end{pmatrix}\,.
\ee
If all $n_i$ are even, the structure factor in the vicinity of the reciprocal lattice point probes only the total magnetization.
If only one $n_i$ is even, it probes $\vec{v}\cdot\vec{B}$, and in the two other cases, there are contributions from both $M(\vec{q})$ and $\vec{v}\cdot\vec{B}(\vec{q})$.

\section{Other results}\label{app:other}
We also provide here the structure factor for the nearest neighbour model. In the notation of \textcite{IsakovGregor04}, where $c_{ab} = \cos\left(\frac{q_a+q_b}{4}\right)$ and $c_{\overline{ab}} = \cos\left(\frac{q_a-q_b}{4}\right)$ with
$Q = c_{xy}^2 +c_{\overline{xy}}^2+c_{yz}^2+c_{\overline{yz}}^2+c_{xz}^2+c_{\overline{xz}}^2-3$ and further defining $s^2_{a} \equiv \sin^2\left(\frac{q_a}{4}\right)$ and $c_{(ab)}\equiv c_{ab}+c_{\overline{ab}}$, the static structure factor is

\begin{widetext}
\be\label{eq:nnexactstructure}
S(\vec{q})_{\rm nn} = \frac{1}{\lambda}\left[2+
\frac{(2-c_{(xy)}-c_{(yz)}-c_{(zy)})t\lambda + \frac{1}{2}(t\lambda)^2-4\left[
      c_{(xy)}s^2_z+ c_{(yz)}s^2_x + c_{(zx)}s^2_y
                  \right]}{3-Q + 2t\lambda + \frac{1}{4}(t\lambda)^2} \right];
\ee
\end{widetext}
where $t=T/J_1$ and $\lambda$ is the solution of \refeq{eq:scgacondition}. At $T \ll J_1$, the zero temperature solution $\lambda = 3/2$ is a good approximation (see Fig.~\ref{fig:scgasolution}). Up to an overall rescaling due to the choice of spin length, \refeq{eq:nnexactstructure} reduces in the limit $t\rightarrow0$ to the zero temperature form provided by \textcite{IsakovGregor04}.

%

\end{document}